# Algorithms for Runtime Generation of Homogeneous Classes of Objects

D. O. Terletskyi

Taras Shevchenko National University of Kyiv, Kyiv/Ukraine, dmytro.terletskyi@gmail.com

*Abstract* – **This paper contains analysis of main modern approaches to dynamic code generation, in particular creation of new classes during program execution. The main attention was paid to universal exploiters of homogeneous classes of objects, which were proposed as a part of such knowledge-representation model as object-oriented dynamic networks, as the tools for creation of new classes of objects in program runtime. As the result, algorithms for implementation of such universal exploiters of classes of objects as union, intersection, difference and symmetric difference were developed. These algorithms can be used in knowledge-based intelligent systems, which are based on object-oriented dynamic networks, and they can be adapted for some object-oriented programming languages with powerful metaprogramming opportunities.**

*Keywords* – **runtime code generation, runtime class generation, universal exploiters of classes, homogeneous classes.**

## I. Introduction

As the result of intensive development of programming languages and technologies during a few last decades, many new programming techniques, tools, technologies and directions within the area have aroused. One of the important and attractive directions within the modern programming is metaprogramming, the main ideas of which is an ability of programs to analyze, modify and generate codes of other programs, including their own. Such approach is aimed at automation of some phases of software development and increasing of adaptability and scalability of the developed software.

Currently code generation is the most interesting part of metaprogramming. It can be used for generation of some parts of programs codes, as well as for generation of whole programs. For today there are two main approaches to code generation: *compile-time code generation (CTCG)* and *runtime code generation (RTCG)* [1, 2].

During CTCG, code generation is performed on the stage of program compilation, when a compiler analyzed meta-structure of a program and transforms its code to corresponding executable machine codes. After that meta-structure of the program is not accessible within the run-time, that is why such approach also known as *static metaprogramming*. Usually it can be implemented within compiled high-level programming languages with static typing, such as C++, C#, Java, Scala, etc.

During RTCG, code generation performs on the stages of program execution, when interpreter can modify existed program' codes and generate new codes. In this case, whole program' meta-structure is accessible for interpreter in runtime, that is why such approach is also known as *dynamic metaprogramming*. Usually it can be implemented within interpreted (and some compiled) high-level programming languages with dynamic typing, such as Smalltalk-80, Squeak, Lisp, Python, Ruby, Groovy, etc.

## II. Code Generation Within Object-Oriented Programming

Nowadays object-oriented programming (OOP) is the most popular and widespread programming paradigm within the area of software development. Many of modern programming languages support OOP that is why large percent of modern software has object-oriented nature.

Taking into account that one of the main concepts of class-based OOP is a class, consequently, code generation process, in most cases, means generation of new classes. It can be achieved using different OOP languages and both mentioned approaches to code generation, however RTCG is more flexible then CTCG.

One of the OOP languages, which support RTCG, is Python, which provides such syntactic construction as metaclasses, metaattributes and metamethods, which allow modification of existed classes of objects, their attributes and methods, and creation of new ones [3-5]. Another powerful OOP language, which supports RTCG, is Ruby, which, similarly to Python, supports mechanisms of reflection and provides ability to change the structure of the classes, their attributes and methods dynamically [6-8]. These languages provide developers with very powerful toolkits for object-oriented RTCG. Using them, a developer can create new classes of objects and manipulate them in runtime in various ways, for example, using for this software creational patterns, polymorphic metaclasses, metaattributes and metamethods, etc.

## III. Class Generation Within Object-Oriented Knowledge Representation

*Runtime code generation* or *runtime class generation* is important not only within area of OOP, it also plays significant role within area of intelligent systems, in particular object-oriented knowledge-based systems (OOKBS) [9]. As it is known, OOKBSs very often operate with models of different essences from various domains. Usually classes are used for modeling of abstract essences, while objects of these classes are used for modeling of concrete essences. For adaptability and scalability of such systems, they must have an ability to create models of new discovered essences within particular domain. Therefore, generation of new classes is an important task for OOKBSs.

However, generation of new classes, using some templates and polymorphic structures, is not enough for intelligent systems, because such systems should also have some analytical abilities, for example an ability to compare a few different classes and find their common and unique parts. Such skills can be very useful in the processes of recognition, classification, learning, decision making, generation and extraction of new knowledge from previously known ones, etc. Such abilities can be implemented using appropriate metaprogramming toolkits, which provide modern OOP languages like Python or Ruby.

## IV. CONCEPTS OF CLASSES WITHIN OBJECT-ORIENTED DYNAMIC NETWORKS

The design and development of any object-oriented KBS requires choosing of particular object-oriented knowledge representation model (OOKRM). Nowadays most famous OOKRMs are frames [10-12], class-based OOP [13, 15] and prototype-based OOP [13, 14]. However, there is another one object-oriented KRM, which called object-oriented dynamic networks (OODN) [16, 9]. All these KRMs are object-oriented ones, but, in the same time, they use different concepts of class and object. Therefore, processes of RTCG within OOKBSs, which are based on these KRMs, will have the differences.

Concept and structure of the class within frames and class-based OOP is very similar, while structure of the class within OODN has some specific peculiarities. First of all, within the frames, as well as within the class-based OOP, there is one kind of classes – homogeneous classes. Objects of such classes have the same structure as their classes. However, within the OODN there are three kinds of class: *homogeneous classes*, *single-core and multi-core heterogeneous classes of objects*. As it was shown in [18, 17, 9], heterogeneous classes have strong connection with homogeneous classes and in some cases are much effective then the last ones.

Let us consider the main definitions.

**Definition 1.** *Homogeneous class of objects $T$ is a tuple of the following form*

$$T = (P(T), F(T)) =$$
$$= ((p_1(T),...,p_n(T)), (f_1(T),...,f_m(T))),$$

*where $P(T)$ is a specification (a vector of properties), which defines some quantity of objects with the same structure, and $F(T)$ is a signature (a vector of methods), which can be applied to them.*

This definition is also suitable for concepts of classes within the frames and class-based OOP. All details about definitions of specifications, signatures, as well as about properties and methods within the OODN, are represented in [9, 16, 18].

**Definition 2.** *Single-core heterogeneous class of objects $T$ is a tuple of the following form*

$$T = (Core(T), pr_1(t_1),..., pr_l(t_l)),$$

*where $Core(T) = (P(T), F(T))$ is a core of the class $T$, which contains properties and methods that are common for types of objects $t_1,...,t_l$, and $pr_i(t_i) = (P(t_i), F(t_i))$, where $i = \overline{1,r}$, $r \leq l$, are their projections, which contain properties and methods which are common only for type $t_i$.*

All details about equivalence of properties and methods within the OOND are represented in [9]. Main peculiarities of single-core heterogeneous classes of objects and their properties are described in [9, 17].

The concept of single-core heterogeneous class of objects shows the difference between notion of *class* and *type*, which are equivalent within the frames and class-based OOP. Analyzing Def. 2, we can see that single-core heterogeneous class of objects can define objects of different structure, i.e. objects of different types. These types are not equivalent, but can have some equivalent properties or methods. That is why within OODN notion of class and of type are different.

**Definition 3.** *Type $t_i$ of single-core heterogeneous class of objects $T$ is a homogeneous class of objects $t_i = (Core(T), pr_j(t_j))$, where $Core(T)$ is a core of class $T$, and $pr_j(t_j)$ is its $j$-th projection, where $i = \overline{1,n}$, $j = \overline{1,m}$, $m \leq n$, where $n$ is a quantity of types which are defined by class $T$.*

## V. CLASS GENERATION WITHIN OBJECT-ORIENTED DYNAMIC NETWORKS

One more distinctive feature of OODN is that it provides tools for modification of previously defined and for generation of new classes of objects, called modifiers and exploiters respectively [9, 17]. Let us consider notion of exploiters within the OODN and its connection with RTCG.

General definition of exploiters can be formulated in the following way.

**Definition 4.** *Exploiter is a function (method), which uses objects and classes of objects as unchangeable parameters for creation of new objects, classes, sets and multisets of objects.*

Analyzing this definition, we can conclude that exploiters can be used not only for creation of new classes of objects. However, in this paper we are going to consider their application only for this purpose.

The notion of exploiter allows definition of various exploiters within OODN, however most of them will be locally closed i.e. they cannot be applied to different classes. Nevertheless, there are universal exploiters, which can be applied to any class of objects. Therefore, such universal exploiters of classes, as union, homogeneous intersection, inhomogeneous intersection, difference, symmetric difference and cloning were proposed in [9].

Let us consider definitions of union, homogeneous intersection, difference and symmetric difference exploiters of classes.

**Definition 5.** Union $T_1 \cup \ldots \cup T_n$ of classes of objects $T_1, \ldots, T_n$, $n \geq 2$, which define $l_1, \ldots l_n$ types of objects respectively, where $l_1 \geq 1, \ldots, l_n \geq 1$, is the new class of objects $T_{1\ldots m}$ which defines types of objects $t_1, \ldots, t_m$, such that

$$\forall (t_{w_1}, t_{w_2}), w_1 \neq w_2 \mid Eq(t_{w_1}, t_{w_2}) = 0,$$

where $w_1, w_2 = \overline{1, m}$, $1 \leq m \leq l_1 + \ldots + l_n$, and

$$\left(\forall t_i^k, \exists! t_j^{1\ldots m}\right) \wedge \left(\forall t_j^{1\ldots m}, \exists t_j^k\right) \mid Eq\left(t_j^k, t_j^{1\ldots m}\right) = 1,$$

where $t_j^k$ is a $i$-th type of class $T_k$, where $i = \overline{1, l_k}$, $k = \overline{1, n}$, and $t_j^{1\ldots m}$ is a $j$-th type of class $T_{1\ldots m}$, $j = \overline{1, m}$.

Universal exploiter of union allows creation of new class of objects $T_{1\ldots m}$, which can be homogeneous or heterogeneous, depending on equivalence and level of heterogeneity of classes $T_1, \ldots, T_n$.

**Definition 6.** Homogeneous intersection $T_1 \cap \ldots \cap T_n$ of classes of objects $T_1, \ldots, T_n$, $n \geq 2$, which define $l_1, \ldots, l_n$ types of objects respectively, $l_1 \geq 1, \ldots, l_n \geq 1$, is the new homogeneous class of objects $T$, which defines type of objects $t$, such that

$$(\forall t_i, t \subseteq t_i) \wedge (\neg \exists t' \mid (t \subseteq t') \wedge (t' \subseteq t_i)),$$

where $t_i$ is a type of objects defined by class $T_i$, $i = \overline{1, n}$. Homogeneous intersection of classes of objects $T_1, \ldots, T_n$ exists if, and only if

$$\exists (p_{i_1}(t_1), \ldots, p_{i_n}(t_n)) \mid Eq(p_{i_1}(t_1), \ldots, p_{i_n}(t_n)) = 1,$$

where $p_{i_k}(t_k)$ is $i_k$-th property of type $t_k$, where $i_k = \overline{1, D(t_k)}$ and $k = \overline{1, n}$.

Universal exploiter of homogeneous intersection allows creation of new homogeneous classes of objects $T$, when there are equivalent properties and (or) methods for all classes $T_1, \ldots, T_n$.

**Definition 7.** Difference $T_1 \setminus T_2$ between classes of objects $T_1$ and $T_2$, which define types of objects $t_1^1, \ldots, t_n^1$ and $t_1^2, \ldots, t_m^2$, $n, m \geq 1$, respectively, is the new class of objects $T_{1\setminus 2}$, which defines types of objects $t_1^{1\setminus 2}, \ldots, t_k^{1\setminus 2}$, such that $k \leq n + m$ and

$$\forall \left(t_i^{1\setminus 2}, t_w^2\right), \exists t_j^1 \mid \left(t_i^{1\setminus 2} \subset t_j^1\right) \wedge \neg \exists \left(t_i^{1\setminus 2} \cap t_w^2\right) \wedge$$
$$\wedge \left(\neg \exists t^{1\setminus 2} \mid \left(t_i^{1\setminus 2} \subset t^{1\setminus 2}\right) \wedge \left(t^{1\setminus 2} \subseteq t_j^1\right) \wedge \neg \exists \left(t^{1\setminus 2} \cap t_w^2\right)\right),$$

where $i = \overline{1, k}$, $j = \overline{1, n}$, $w = \overline{1, m}$. The difference between classes of objects $T_1$ and $T_2$ exists if, and only if

$$\exists p_{i_1}\left(t_j^1\right), \exists p_{i_2}\left(t_w^2\right) \mid Eq\left(p_{i_1}\left(t_j^1\right), p_{i_2}\left(t_w^2\right)\right) = 0,$$

where $p_{i_1}\left(t_j^1\right)$ is $i_1$-th property of type $t_j^1$, $i_1 = \overline{1, D(t_j^1)}$, and $p_{i_2}\left(t_w^2\right)$ is $i_2$-th property of type $t_w^2$, $i_2 = \overline{1, D(t_w^2)}$.

Universal exploiter of difference allows creation of new class of objects $T_{1\setminus 2}$, which can be homogeneous or heterogeneous, depending on level of heterogeneity of classes $T_1$ and $T_2$, when class $T_1$ has unique properties and (or) methods.

**Definition 8.** Symmetric difference $T_1 \div T_2$ between classes of objects $T_1$ and $T_2$, which define types of objects $t_1^1, \ldots, t_n^1$ and $t_1^2, \ldots, t_m^2$, $n, m \geq 1$, respectively is the new heterogeneous class of objects $T_{1\div 2}$, which defines types of objects $t_1^{1\setminus 2}, \ldots, t_w^{1\setminus 2}$ and $t_1^{2\setminus 1}, \ldots, t_q^{2\setminus 1}$, such that $w + q \leq n + m$. Symmetric difference between classes of objects $T_1$ and $T_2$ exists if, and only if

$$\exists p_{i_1}\left(t_i^1\right), \exists p_{i_2}\left(t_j^2\right) \mid Eq\left(p_{i_1}\left(t_i^1\right), p_{i_2}\left(t_j^2\right)\right) = 0,$$

where $p_{i_1}\left(t_i^1\right)$ is an $i_1$-th property of type $t_i^1$, $i = \overline{1, D(t_i^1)}$, $i = \overline{1, n}$, and $p_{i_2}\left(t_j^2\right)$ is an $i_2$-th property of type $t_j^2$, $i_2 = \overline{1, D(t_j^2)}$, $j = \overline{1, m}$.

Universal exploiter of symmetric difference allows creation of new class of objects $T_{1\div 2}$, which can be homogeneous or heterogeneous, depending on level of heterogeneity of classes $T_1$ and $T_2$, when they have unique properties and (or) methods.

All these universal exploiters are formally defined, and can be extend for the classes within class-based OOP. However, it is necessary to develop corresponding efficient algorithms for their practical implementation.

## VI. ALGORITHMS FOR IMPLEMENTATION OF SOME UNIVERSAL EXPLOITERS OF CLASSES

Analyzing definition of union' exploiter, we can conclude that union of $n > 2$ classes of objects requires checking of equivalence for all elements of all possible $n$-tuples of properties $(p_1 \subset P(t_1), p_2 \subset P(t_2), \ldots, p_n \subset P(t_n))$ and methods $(f_1 \subset F(t_1), f_2 \subset F(t_2), \ldots, f_n \subset F(t_n))$ of these classes. Therefore, if there is no any specific information about structures of these classes, which could help to reduce the number of such tuples, the approximate complexity of the algorithm is equal to

$$D(t_1) \times D(t_2) \times \ldots \times D(t_n) +$$
$$+ func(t_1) \times func(t_2) \times \ldots \times func(t_n),$$

where $D(t_i)$ is a dimension of the class $t_i$, $i = \overline{1, n}$, i.e. quantity of properties, and $func(t_i)$ is a functionality of the class $t_i$, i.e. quantity of methods. Taking into account this fact and Def. 5, it is possible to propose the following algorithm for union of $n \geq 2$ classes.

**Algorithm of union**

**Input:** classes of objects $t_1,...,t_n$ (or their copies).

1. Consider and check the equivalence of all elements of all possible $n$-tuples $(p_1 \subset P(t_1),..., p_n \subset P(t_n))$ and $(f_1 \subset F(t_1),..., f_n \subset F(t_n))$ constructed from the classes $t_1,...,t_n$.

2. If at some iteration such equivalence is found:
   a Copy this property (method) to the core of new class of objects;
   b Delete this property (method) from classes of objects $t_1,...,t_n$ (future projections).

3. Repeat steps 1 and 2 until the end of consideration and comparison of all possible $n$-tuples of properties (methods) of classes $t_1,...,t_n$.

**Output:** new single-core heterogeneous class of objects.

As we can see, the algorithm can receive as the input parameters classes of objects or their copies. If it receives access to classes, then they will be transformed into the parts of new class of objects. That is why if we need these classes to be unchanged after creation of their union, then we should use their copies.

Analyzing Def. 5 and Def. 6, we can conclude that intersection of classes $t_1,...,t_n$ can be computed during the calculation of their union, in this case core of the obtained class is the intersection of classes $t_1,...,t_n$. Taking into account Def. 6, it is possible to propose the following algorithm for intersection of $n \geq 2$ classes.

**Algorithm of intersection**

**Input:** classes of objects $t_1,...,t_n$.

1. Consider and check the equivalence of all elements of all possible $n$-tuples $(p_1 \subset P(t_1),..., p_n \subset P(t_n))$ and $(f_1 \subset F(t_1),..., f_n \subset F(t_n))$ constructed from the classes $t_1,...,t_n$.

2. If at some iteration such equivalence is found:
   a Copy this property (method) to new class of objects.

3. Repeat steps 1 and 2 until the end of consideration and comparison of all possible $n$-tuples of properties (methods) of classes $t_1,...,t_n$.

**Output:** new homogeneous class of objects if intersection among classes $t_1,...,t_n$ exists.

Analyzing Def. 7 we can see that it defines intersection of classes as binary operation, however it can be generalized for the case of n classes. Taking into account Def. 7, it is possible to propose the following algorithm for difference between class $t$ and $n \geq 2$ classes.

**Algorithm of difference**

**Input:** class of objects $t$ (or its copy) and classes of objects $t_1,...,t_n$.

1. Consider and check the equivalence of all elements of all possible tuples of properties $(p \subset P(t), p_i \subset P(t_i))$, and methods $(f \subset F(t), f_i \subset F(t_i))$, constructed from the classes $t, t_1,...,t_n$, where $i = \overline{1, n}$.

2. If at some iteration such equivalence is found:
   a Delete this property (method) from the class $t$.

3. Repeat steps 1 and 2 until the end of consideration and comparison of all possible tuples of properties (methods) of classes $t, t_1,...,t_n$.

**Output:** new homogeneous class of objects if difference between class $t$ and classes $t_1,...,t_n$ exists.

Similarly to union, the algorithm can receive as one of the input parameters class $t$ or its copy. If it receives access to class, then it will be transformed into the new class of objects. That is why if we need this class to be unchanged after creation of the difference between it and classes $t_1,...,t_n$, then we should use the copy of $t$.

Taking into account Def. 8, it is possible to propose the following algorithm for symmetric difference between classes $t_1$ and $t_2$.

**Algorithm of symmetric difference**

**Input:** classes of objects $t_1$ and $t_2$ (or their copies).

1. Consider and check the equivalence of all elements of all possible tuples $(p_1 \subset P(t_1), p_2 \subset P(t_2))$ and $(f_1 \subset F(t_1), f_2 \subset F(t_2))$ constructed from the classes $t_1$ and $t_2$.

2. If at some iteration such equivalence is found:
   a Delete this property (method) from classes $t_1$ and $t_2$.

3. Repeat steps 1 and 2 until the end of consideration and comparison of all possible tuples of properties (methods) of classes $t_1$ and $t_2$.

**Output:** new single-core inhomogeneous class of objects if symmetric difference between classes $t_1$ and $t_2$ exists.

Similarly to union and difference, the algorithm can receive as the input parameters classes of objects or their copies. If it receives access to classes, then they will be transformed into the parts of new class of objects. That is why if we need these classes to be unchanged after creation of their symmetric difference, then we should use their copies.

## VII. Conclusions

An ability of knowledge-based intelligent systems to generate new classes of objects in runtime is very important feature, which allow increasing of adaptability and scalability of such systems. Therefore, universal exploiters of homogeneous classes of objects were considered in the paper as a tool for generation of new classes of objects in program runtime.

The main result of the paper is the algorithms for implementation of universal exploiters of classes of objects, which allow dynamic creation of new classes. As the result, algorithms for union, intersection, difference and symmetric difference of classes of objects were proposed. These algorithms can be useful within the knowledge-based intelligent systems, which are based on object-oriented dynamic networks, and also can be adapted for some high-level object-oriented programming languages equipped by powerful metaprogramming toolkits, such as Python, Ruby, etc.

However, despite all noted advantages, proposed algorithms require further analysis and optimization.